\documentclass[preprint,aps,prd,nofootinbib]{revtex4}
\usepackage{graphicx}
\usepackage{subfigure}

\newcommand{\nn}{\nonumber}

\def\/{\over}

\begin{document}

\title{Thermal nature of de Sitter spacetime \\ and spontaneous excitation of atoms }

\author{ Zhiying Zhu and  Hongwei Yu
}
\affiliation{ Department of Physics and Institute of  Physics,
Hunan Normal University, Changsha, Hunan 410081, China  }

\begin{abstract}
We consider, in de Sitter spacetime,  both freely falling and static
two-level atoms in interaction with a conformally coupled massless
scalar field in the de Sitter-invariant vacuum, and separately
calculate the contributions of vacuum fluctuations and radiation
reaction to the atom's spontaneous excitation rate.  We find that
spontaneous excitations occur even for the freely falling atom as if
there is a thermal bath of radiation at the Gibbons-Hawking
temperature and we thus recover, in a different physical context,
the results of Gibbons and Hawking that reveals the thermal nature
of de Sitter spacetime. Similarly, for the case of the static atom,
our results show that the atom also perceives a thermal bath which
now arises as a result of the intrinsic thermal nature of de Sitter
spacetime and the Unruh effect associated with the inherent
acceleration of the atom.

\end{abstract}

\maketitle

\baselineskip=16pt

\section{introduction}

Spontaneous emission is one of the most interesting problems in the
interaction of atoms with quantum fields and so far mechanisms such
as vacuum fluctuations \cite{Welton48, CPP83}, radiation reaction
\cite{Ackerhalt73}, or a combination of them \cite{Milonni88} have
been put forward to explain why spontaneous emission occurs. The
controversy arises because of the freedom in choices of ordering of
commuting operators of atom and field in a Heisenberg picture
approach to the problem and  was resolved when Dalibard, Dupont-Roc
and CohenTannoudji(DDC) argued in Ref.\cite{Dalibard82} and
Ref.\cite{Dalibard84} that there exists a symmetric operator
ordering that renders the distinct contributions of vacuum
fluctuations and radiation reaction to the rate of  change of an
atomic observable separately Hermitian. If one demands such an
ordering, each contribution can possess an independent physical
meaning. The DDC formalism resolves the problem of stability for
ground-state atoms when only radiation reaction is considered and
the problem of ``spontaneous absorption" of atoms in vacuum when
only vacuum fluctuations are taken into account. Using this
formalism one can show that for ground-state atoms, the
contributions of vacuum fluctuations and radiation reaction to the
rate of change of the mean excitation energy cancel exactly and this
cancellation forbids any transitions from the ground state and thus
ensures atom's stability. While for any initial excited state,
 the rate of change of atomic energy acquires equal
contributions from vacuum fluctuations and from radiation reaction.

Recently, there has been a great deal of interest in the application
of the DDC formalism to accelerated atoms and those in the
background of a black hole~\cite{Audretsch94,H. Yu05,ZYL06,H.
Yu06,ZY07,zwt yu2,zwt yu3}. These studies reveal intriguing
relationships between spontaneous excitation of an atom and the
Unruh effect as well as the Hawking radiation. In a flat spacetime,
the spontaneous excitation of a uniformly accelerated atom in
interaction with vacuum scalar \cite{Audretsch94,H. Yu05,ZY07} and
electromagnetic fields \cite{ZYL06,H. Yu06} has been studied. It is
found that for a ground state atom in uniformly accelerated motion
through the Minkowski vacuum, there is no longer perfect balance
between vacuum fluctuations and radiation reaction. As a result, the
spontaneous excitation rate of the atom is nonzero and furthermore
the rate is exactly what one would obtain
 assuming  the existence of a thermal bath at the Unruh
 temperature.
Inspired by an equivalence principle-type argument,
 the spontaneous excitation rate of atoms in interaction
with a massless scalar field in an interesting kind of curved
spacetimes, i.e., the curved background of a black hole, has
recently been studied~\cite{zwt yu2,zwt yu3}, in both
 the Hartle-Hawking vacuum and Unruh vacuum.
 The results obtained may be
considered as providing a different approach to derivation of the
Hawking effect, since they show that a static atom in the exterior
of a black hole would spontaneously excite as if immersed in a
thermal bath of Hawking radiation.

As natural step forward, we are interested, in the present paper, in
the spontaneous excitation of atoms in yet another kind of special
curved spacetime--de Sitter spacetime.  De Sitter spacetime, being
maximally symmetric, enjoys an important status among the curved
spacetimes similar to that of Minkowski spacetime,  and more
importantly, it has attracted  a surge of renewed interest  in
recent years for the following reason: First, recent observations,
together with the theory of inflation, suggest that our universe may
approach de Sitter geometries in both the far past and the far
future, and second, there may exist a holographic duality between
quantum gravity on de Sitter spacetime and a conformal field theory
living on the boundary identified with the timelike infinity of de
Sitter spacetime~\cite{DSCFT}. Therefore,  it is certainly of
interest to examine the spontaneous excitation of atoms in this
spacetime and this is what we plan to do in the present paper. Using
the DDC formalism, we will calculate the spontaneous excitation rate
of both a freely falling atom  and a static one with an inherent
acceleration in interaction with vacuum fluctuations of quantized
massless conformally coupled scalar fields in de Sitter spacetime.
Let us note that
 the quantization of  scalar fields in this spacetime has been extensively
studied in the literature \cite{QFT,Bunch and Davies,Tagirov,Ford,
Schomblond,Mottola,Allen,Allen87,Polarski,Polarski prd}.

When vacuum fluctuations are concerned in a curved spacetime, one
first has to specify the vacuum states.  The vacuum states in de
Sitter spacetime can be classified into two categories: one is the
de Sitter-invariant states, the other states are those which break
de Sitter invariance \cite{Allen}. Generally, the de
Sitter-invariant vacuum, whose status in de Sitter space is just
like Minkowski vacuum in the flat space, is deemed to be a natural
vacuum. So, we will investigate the spontaneous excitation of atoms
in interaction with a conformally coupled massless scalar field in
the de Sitter-invariant vacuum. We will show that for an atom moving
on a timelike geodesic (freely falling), the spontaneous excitation
rate is what one would expect if the atom  were in a thermal bath of
radiation at the Gibbons-Hawking temperature~\cite{Gibbons}. While
for a static atom in the de Sitter-invariant vacuum, we find that it
also may spontaneously excite as if immersed in a thermal bath at a
temperature equal to the square root of the sum of the squared
Gibbons-Hawking temperature and the squared Unruh temperature
associated with the inherent acceleration of the atom.

It is worth pointing out that the difference between analyzing the
spontaneous excitation of atoms using the DDC formalism as we do in
the current paper and similar previous calculations of the response
of model detectors in de Sitter spacetime~\cite{QFT,Gibbons} lies in
that our discussions provide a physically appealing interpretation
of the thermal response of the detector, i.e.,  a transparent
illustration for why the detector clicks, since the spontaneous
excitation of the atoms can be considered as the actual physical
process that is actually taking place inside a model detector
revealing the thermal nature of de Sitter spacetime.

\section{The general formalism}

We consider a pointlike two-level atom in interaction with a
conformally coupled massless scalar field in de Sitter spacetime and
assume that the atom has a stationary trajectory $x(\tau)$, where
$\tau$ denotes the proper time on the trajectory. This stationary
trajectory guarantees that the atom has stationary states,
$|-\rangle$ and $|+\rangle$, with energy $-{1\/2}\omega_{0}$ and
${1\/2}\omega_{0}$. The atom's Hamiltonian with respect to its
proper time $\tau$ can be written as~\cite{Dicke}
\begin{eqnarray}
H_A(\tau)=\omega_0R_3(\tau)\;,
\end{eqnarray}
where $R_3(0)={1\/2}|+\rangle\langle+|-{1\/2}|-\rangle\langle-|$.
The free Hamiltonian of the quantum scalar field is
\begin{eqnarray}
H_F(\tau)=\int d^3k~\omega_{\vec{k}}~a_{\vec{k}}^\dag~
a_{\vec{k}}{dt\/d\tau}\;,
\end{eqnarray}
where $a_{\vec{k}}^\dag$ and $a_{\vec{k}}$ denote the creation and
annihilation operators with momentum $\vec{k}$. The Hamiltonian that
describes the interaction between the atom and the quantum field is
given by~\cite{Audretsch94}
\begin{eqnarray}
H_I(\tau)=\mu R_2(\tau)\phi(x(\tau))\;.
\end{eqnarray}
Here $\mu$ is a coupling constant which we assume to be small,
$R_2(0)={1\/2}i[R_-(0)-R_+(0)]$, where $R_+(0)=|+\rangle\langle-|$
and $R_-(0)=|-\rangle\langle+|$. $\phi(x)$ is the scalar field
operator in de Sitter spacetime and it satisfies the wave equation
\begin{eqnarray}
(\nabla_\mu \nabla^\mu+m^2+\xi R)\phi=0\;,\label{w eq}
\end{eqnarray}
where $m$ is the mass of the scalar field,  $\xi$ is a coupling
constant and $R$ is the scalar curvature. The coupling is effective
only on the trajectory  of the atom.

Then we can write down the Heisenberg equations of motion for the
dynamical variables of the atom and field from the Hamiltonian
$H=H_A+H_F+H_I$. The solutions of the equations of motion can be
split into the two parts: a free part, which is present even in the
absence of the coupling, and a source part, which is caused by the
interaction of the atom and field. We assume that the initial state
of the field is the de Sitter-invariant vacuum (also known as
Euclidean or Bunch-Davies vacuum \cite{Bunch and Davies})
$|0\rangle$, and the atom is prepared in the state $|a\rangle$,
which may be $|+\rangle$ or $|-\rangle$. Choosing a symmetric
ordering between the atom and field variables, we can separate the
two contributions of vacuum fluctuations and radiation reaction to
the rate of change of $\langle H_A\rangle$ (~cf.
Refs.~\cite{Audretsch94,Dalibard82,Dalibard84}~),
\begin{eqnarray}
\biggl\langle{dH_A(\tau)\/d\tau}\biggr\rangle_{VF}=2i\mu^2\int_{\tau_0}^\tau
d\tau'C^F(x,x'){d\/d\tau}\chi^A(\tau,\tau')\;,\label{vf}
\end{eqnarray}
\begin{eqnarray}
\biggl\langle{dH_A(\tau)\/d\tau}\biggr\rangle_{RR}=2i\mu^2\int_{\tau_0}^\tau
d\tau'\chi^F(x,x'){d\/d\tau}C^A(\tau,\tau')\;,\label{rr}
\end{eqnarray}
where $| \rangle = |a,0 \rangle$ represents the atom in the state
$|a\rangle$ and the field in the de Sitter-invariant vacuum state
$|0 \rangle$. They are expressed in terms of the statistical
functions of the free part of the atom's variable, $R_2^f$
\begin{eqnarray}
C^A(\tau,\tau')={1\/2}\langle
a|\{R_2^f(\tau),R_2^f(\tau')\}|a\rangle\;,
\end{eqnarray}
\begin{eqnarray}
\chi^A(\tau,\tau')={1\/2}\langle
a|~[R_2^f(\tau),R_2^f(\tau')]~|a\rangle\;,
\end{eqnarray}
and those of the field's, $\phi^f$,
\begin{eqnarray}
C^F(x(\tau),x(\tau'))={1\/2}\langle0|\{\phi^f(x(\tau)),\phi^f(x(\tau'))\}|0\rangle
\;,\label{cf}
\end{eqnarray}
\begin{eqnarray}
\chi^F(x(\tau),x(\tau')={1\/2}\langle0|[\phi^f(x(\tau)),\phi^f(x(\tau'))]|0\rangle
\;.\label{xf}
\end{eqnarray}
$C^F$ ($C^A$) is called the symmetric correlation function of the
field (atom), $\chi^F$ ($\chi^A$) its linear susceptibility. The
explicit forms of the statistical functions of the atom are given by
\begin{eqnarray}
C^A(\tau,\tau')={1\/2}\sum_b|\langle a|R_2^f(0)|b\rangle|^2\biggl(
e^{i\omega_{ab}(\tau-\tau')}+e^{-i\omega_{ab}(\tau-\tau')}\biggr)\;,\label{ca}
\end{eqnarray}
\begin{eqnarray}
\chi^A(\tau,\tau')={1\/2}\sum_b|\langle
a|R_2^f(0)|b\rangle|^2\biggl(
e^{i\omega_{ab}(\tau-\tau')}-e^{-i\omega_{ab}(\tau-\tau')}\biggr)\;,\label{xa}
\end{eqnarray}
where $\omega_{ab}=\omega_a-\omega_b$ and the sum extends over a
complete set of atomic states.

\section{ Spontaneous  excitation of a freely falling atom in de Sitter spacetime}

In this Section we will consider a freely moving atom interacting
with a conformally coupled massless scalar field in de Sitter
spacetime. As is well known, different coordinates systems can be
used to parameterize de Sitter spacetime \cite{QFT}. The rate of
change of the atomic energy is a scalar and should be independent of
the coordinates. Here we choose to work in the global coordinate
system, in which the line element is expressed as
\begin{eqnarray}
ds^2=dt^2-\alpha^2\cosh^2(t/\alpha)[d\chi^2+\sin^2\chi(d\theta^2+\sin^2\theta
d\varphi^2)]\;.\label{ds1}
\end{eqnarray}
Here $\alpha=3^{1/2}\Lambda^{-1/2}$, where $\Lambda$ is the
cosmological constant, and the scalar curvature $R=12\alpha^{-2}$.
The canonical quantization of a massive scalar field with this
metric has been dealt with in
Refs.~\cite{QFT,Tagirov,Mottola,Allen,Allen87}. In coordinates
(\ref{ds1}), the wave equation (\ref{w eq}) for a massive scalar
field becomes
\begin{eqnarray}
\biggl[{1\/\cosh^3t/\alpha}{\partial\/\partial
t}\biggl({\cosh^3{t\/\alpha}}~{\partial\/\partial
t}\biggr)-{\Delta\/\alpha^2\cosh^2t/\alpha}+m^2+\xi
R\biggr]\phi=0\;,\label{w eq1}
\end{eqnarray}
where the Laplacian
\begin{eqnarray}
\Delta={1\/\sin^2\chi}\biggl[{\partial\/\partial\chi}\biggl(\sin^2\chi{\partial\/\partial\chi}\biggr)
+{1\/\sin\theta}{\partial\/\partial\theta}\biggl(\sin\theta{\partial\/\partial\theta}\biggr)
+{1\/\sin^2\theta}{\partial^2\/\partial\varphi^2}\biggr]\;.
\end{eqnarray}
From (\ref{w eq1}) one can get the eigenmodes, and define a de
Sitter-invariant vacuum. Then the Wightman function can be written
as \cite{Allen87}
\begin{eqnarray}
G^+(x(\tau),x(\tau')))=-{1\/16\pi\alpha^2}~{{1\/4}-\nu^2\/\cos
\pi\nu}~F\biggl({3\/2}+\nu,{3\/2}-\nu;2;{1-Z(x,x')\/2}\biggr)\;,
\end{eqnarray}
where $F$ is a hypergeometric function, and
\begin{eqnarray}
&&Z(x,x')=\sinh{t\/\alpha}\sinh{t'\/\alpha}-\cosh{t\/\alpha}\cosh{t'\/\alpha}\cos\Omega\;\nonumber\\&&
\cos\Omega=\cos\chi\cos\chi'+\sin\chi\sin\chi'[\cos\theta\cos\theta'+\sin\theta\sin\theta'\cos(\varphi-\varphi')]'\;,\nonumber\\&&
\nu=\biggl[{9\/4}-{12\/R}(m^2+\xi R)\biggr]^{1/2}\;.
\end{eqnarray}
In the massless, conformally coupled limit, for a freely falling
atom,
the Wightman function becomes
\begin{eqnarray}
G^+(x(\tau),x(\tau')))=-{1\/16\pi^2\alpha^2\sinh^2({\tau-\tau'\/2\alpha}-i\varepsilon)}\;.
\end{eqnarray}
Then the statistical functions of the field, (\ref{cf}) and
(\ref{xf}), can be obtained
\begin{eqnarray}
C^F(x(\tau),x(\tau'))=-{1\/32\pi^2\alpha^2}\biggl[{1\/\sinh^2({\tau-\tau'\/2\alpha}-i\varepsilon)}
+{1\/\sinh^2({\tau-\tau'\/2\alpha}+i\varepsilon)}\biggr]\;,\label{cf1}
\end{eqnarray}
\begin{eqnarray}
\chi^F(x(\tau),x(\tau'))={i\/4\pi
\cos({i(\tau-\tau')\/2\alpha})}\delta'(\tau-\tau')\;.\label{xf1}
\end{eqnarray}
With a substitution $u=\tau-\tau'$, the contributions of vacuum
fluctuations (\ref{vf}) and radiation reaction (\ref{rr}) to the
rate of change of the atomic energy read
\begin{eqnarray}
\biggl\langle{dH_A(\tau)\/d\tau}\biggr\rangle_{VF}=-{\mu^2\/32\pi^2\alpha^2}\sum_b\omega_{ab}|\langle
a|R_2^f(0)|b\rangle|^2\int^{+\infty}_{-\infty}du\biggl[{1\/\sin^2({i
u\/2\alpha}+\varepsilon)}+{1\/\sin^2({i
u\/2\alpha}-\varepsilon)}\biggr]e^{i\omega_{ab}u}\nonumber\\
\end{eqnarray}
and
\begin{eqnarray}
\biggl\langle{dH_A(\tau)\/d\tau}\biggr\rangle_{RR}
=-{i\mu^2\/4\pi}\sum_b\omega_{ab}|\langle
a|R_2^f(0)|b\rangle|^2\int^{+\infty}_{-\infty}du{e^{i\omega_{ab}u}\/\cos({i
u\/2\alpha})}\delta'(u)\;,
\end{eqnarray}
where we have extended the range of integration to infinity for
sufficiently long times $\tau-\tau_0$. With the help of residue
theorem, we can evaluate the integrals to get
\begin{eqnarray}
\biggl\langle{dH_A(\tau)\/d\tau}\biggr\rangle_{VF}&=&-{\mu^2\/4\pi}\biggl[
\sum_{\omega_a>\omega_b} \omega_{ab}^2|\langle
a|R_2^f(0)|b\rangle|^2\biggl(1+{2\/e^{2\pi\alpha\omega_{ab}}-1}\biggr)
\nn\\&&-\sum_{\omega_a<\omega_b}\omega_{ab}^2|\langle
a|R_2^f(0)|b\rangle|^2\biggl(1+{2\/e^{2\pi\alpha|\omega_{ab}|}-1}\biggr)\biggr]\label{vf1}
\end{eqnarray}
for the contributions of vacuum fluctuations to the rate of change
of atomic energy, and
\begin{eqnarray}
\biggl\langle{dH_A(\tau)\/d\tau}\biggr\rangle_{RR}
=-{\mu^2\/4\pi}\biggl(\sum_{\omega_a>\omega_b}\omega_{ab}^2|\langle
a|R_2^f(0)|b\rangle|^2+\sum_{\omega_a<\omega_b}\omega_{ab}^2|\langle
a|R_2^f(0)|b\rangle|^2\biggr)\label{rr1}
\end{eqnarray}
for that of radiation reaction. Comparison with the case of an
inertial atom in the Minkowski vacuum \cite{Audretsch94} shows that
the contribution of radiation reaction to the rate of change of the
atomic energy is the same as that in the flat spacetime, and thus is
independent of the spacetime curvature. But the contribution of
vacuum fluctuations is modified by the appearance of a thermal like
term. Adding up the contributions of the vacuum fluctuations
(\ref{vf1}) and radiation reaction (\ref{rr1}) we arrive at the
total rate of change of the atomic energy:
\begin{eqnarray}
\biggl\langle{dH_A(\tau)\/d\tau}\biggr\rangle_{tot}&=&-{\mu^2\/2\pi}\biggl[
\sum_{\omega_a>\omega_b} \omega_{ab}^2|\langle
a|R_2^f(0)|b\rangle|^2\biggl(1+{1\/e^{2\pi\alpha\omega_{ab}}-1}\biggr)
\nn\\&&-\sum_{\omega_a<\omega_b}\omega_{ab}^2|\langle
a|R_2^f(0)|b\rangle|^2{1\/e^{2\pi\alpha|\omega_{ab}|}-1}\biggr]\;.\label{tot1}
\end{eqnarray}
For an atom in its ground state in the de Sitter-invariant vacuum,
there is a positive contribution. So the atom spontaneously excites,
just as if it were in a bath of blackbody radiation at the
temperature $T=1/2\pi\alpha$, which is exactly the temperature found
by Gibbons and Hawking~\cite{Gibbons} by examining  the response of
a freely moving particle-detector in de Sitter spacetime. We
therefore recover, in a different physical context, the results of
Gibbons and Hawking that reveals the thermal nature of de Sitter
spacetime~\cite{Gibbons}. It should be pointed out that since the
cosmological constant is a very small number, so the temperature $T$
is insignificant in terms of the experimental observation.

\section{Spontaneous excitation of a static atom in de Sitter spacetime }

Now, we will calculate the spontaneous excitation rate of a static
atom in  de Sitter spacetime interacting with a conformally coupled
massless scalar field in the de Sitter-invariant vacuum.  For this
purpose, it is convenient to work in the static coordinate system in
which the line element is written as
\begin{eqnarray}
ds^2=\biggl(1-{r^2\/\alpha^2}\biggr)d\tilde{t}^2-\biggl(1-{r^2\/\alpha^2}\biggr)^{-1}dr^2-r^2(d\theta^2+\sin^2\theta
d\varphi^2)\;.\label{static}
\end{eqnarray}
This metric possesses a coordinate singularity of the type of an
event horizon at $r=\alpha$. The coordinates $(\tilde{t},r,\theta
,\varphi)$ only cover part of de Sitter spacetime, just like the
Rindler wedge. For an atom at rest in the static coordinates system,
its inherent acceleration is
\begin{eqnarray}
a={r\/\alpha^2}\biggl(1-{r^2\/\alpha^2}\biggr)^{-1/2}\;.
\end{eqnarray}
The static coordinates are related to the global coordinates by
\begin{eqnarray}
r=\alpha\cosh(t/\alpha)\sin\chi\;, \ \ \
\tanh(\tilde{t}/\alpha)=\tanh(t/\alpha)\sec\chi\;.
\end{eqnarray}
It should be noted that the worldlines of observers in the global
and static coordinates coincide at $r=0$ and $\chi=0$ and an atom at
rest in the static coordinates with $r\neq0$ will be accelerated
with respect to an atom at rest in the global coordinates with
$\chi=0$. In the static de Sitter metric (\ref{static}), one can
obtain a complete set of mode solutions of (\ref{w eq})
\cite{Polarski,Polarski prd}, and chooses a de Sitter-invariant
vacuum. Then the Wightman function for a massless conformally
coupled scalar field is given by~\cite{Polarski2,Gal'tsov}
\begin{eqnarray}
G^+(x(\tau),x(\tau')))=-{1\/8\pi^2\alpha^2}{\cosh({r^\ast\/\alpha})\cosh({{r^\ast}'\/\alpha})\/
\cosh({\tilde{t}-\tilde{t}'\/\alpha}-i\varepsilon)
-\cosh({r^\ast-{r^\ast}'\/\alpha})}\;,\label{gstatic1}
\end{eqnarray}
where
\begin{eqnarray}
r^\ast={\alpha\/2}\ln{\alpha+r\/\alpha-r}\;.
\end{eqnarray}
For a static atom, Eq.~(\ref{gstatic1}) becomes
\begin{eqnarray}
G^+(x(\tau),x(\tau')))=-{1\/16\pi^2\kappa^2\sinh^2({\tau-\tau'\/2\kappa}-i\varepsilon)}\;,\label{gstatic2}
\end{eqnarray}
where $\kappa=\alpha \sqrt{g_{00}}$, and we have used the definition
\begin{eqnarray}
\Delta\tau=\sqrt{g_{00}}\Delta\tilde{t}\;.
\end{eqnarray}
One can then show that
\begin{eqnarray}
C^F(x(\tau),x(\tau'))=-{1\/32\pi^2\kappa^2}\biggl[{1\/\sinh^2({\tau-\tau'\/2\kappa}-i\varepsilon)}
+{1\/\sinh^2({\tau-\tau'\/2\kappa}+i\varepsilon)}\biggr]\;,
\end{eqnarray}
\begin{eqnarray}
\chi^F(x(\tau),x(\tau'))={i\/4\pi
\cos({i(\tau-\tau')\/2\kappa})}\delta'(\tau-\tau')\;.
\end{eqnarray}
With the statistical functions given, we can compute the
contributions of the vacuum fluctuations and radiation reaction to
the rate of change of the atomic energy to get
\begin{eqnarray}
\biggl\langle{dH_A(\tau)\/d\tau}\biggr\rangle_{VF}=&=&-{\mu^2\/4\pi}\biggl[
\sum_{\omega_a>\omega_b} \omega_{ab}^2|\langle
a|R_2^f(0)|b\rangle|^2\biggl(1+{2\/e^{2\pi\kappa\omega_{ab}}-1}\biggr)
\nn\\&&-\sum_{\omega_a<\omega_b}\omega_{ab}^2|\langle
a|R_2^f(0)|b\rangle|^2\biggl(1+{2\/e^{2\pi\kappa|\omega_{ab}|}-1}\biggr)\biggr].\label{vf2}
\end{eqnarray}
and
\begin{eqnarray}
\biggl\langle{dH_A(\tau)\/d\tau}\biggr\rangle_{RR}
=-{\mu^2\/4\pi}\biggl(\sum_{\omega_a>\omega_b}\omega_{ab}^2|\langle
a|R_2^f(0)|b\rangle|^2+\sum_{\omega_a<\omega_b}\omega_{ab}^2|\langle
a|R_2^f(0)|b\rangle|^2\biggr)\;.\label{rr2}
\end{eqnarray}
Adding up the contributions of the vacuum fluctuations (\ref{vf2})
and radiation reaction (\ref{rr2}) we obtain the total rate of
change of the atomic energy:
\begin{eqnarray}
\biggl\langle{dH_A(\tau)\/d\tau}\biggr\rangle_{tot}&=&-{\mu^2\/2\pi}\biggl[
\sum_{\omega_a>\omega_b} \omega_{ab}^2|\langle
a|R_2^f(0)|b\rangle|^2\biggl(1+{1\/e^{2\pi\kappa\omega_{ab}}-1}\biggr)
\nn\\&&-\sum_{\omega_a<\omega_b}\omega_{ab}^2|\langle
a|R_2^f(0)|b\rangle|^2{1\/e^{2\pi\kappa|\omega_{ab}|}-1}\biggr]\;.\label{tot2}
\end{eqnarray}
The above calculations show that the contribution of radiation
reaction to the spontaneous excitation rate is independent of the
spacetime curvature and the acceleration. From Eq.~(\ref{tot2}) we
can see that for a static atom with an inherent acceleration in its
ground state in the de Sitter-invariant vacuum, there is a positive
contribution to the spontaneous excitation rate. Thus transitions
from ground state to the excited states become possible, and they
occur just as if the atoms were immersed in a thermal bath at the
temperature
\begin{eqnarray}
T={1\/2\pi\kappa}={1\/2\pi\sqrt{g_{00}}\alpha}.
\end{eqnarray}
It is interesting to note that the above temperature can be written
in the form
\begin{eqnarray}
T^2=\biggl({1\/2\pi\alpha}\biggr)^2+\biggl({a\/2\pi}\biggr)^2=
T_{GH}^2+T_{U}^2\;.\label{T^2}
\end{eqnarray}
The first spacetime curvature dependent term in the above equation
is the squared Gibbons-Hawking temperature of de Sitter spacetime,
which is the temperature of thermal radiation as perceived by a
freely falling atom, and the second  acceleration dependent term is
the squared Unruh temperature which arises as a result of the Unruh
effect. Thus, the thermal radiation as felt by a static atom in de
Sitter spacetime is a combination of two different effects. One
arises from the thermal nature of de Sitter spacetime itself and is
characterized by the Gibbons-Hawking temperature and the other from
the acceleration induced thermal effect characterized by the Unruh
temperature. This is quite similar to what happens for a static atom
outside a black hole, where one finds that both the Hawking effect
and the Unruh effect contribute to the bath of thermal radiation
encountered by the atom~\cite{zwt yu2,zwt yu3}, although the
detailed relationship is different.
Relation Eq.~(\ref{T^2}) agrees with the result obtained in other
different physical contexts~\cite{NPT,Deser}.
 Let us note that $T$ diverges as $r$
approaches $\alpha$ as a consequence of the blowup of $T_{U}$. The
reason is that to hold the atom static at the event horizon, an
infinite inherent acceleration is needed.

\section{summary}

We have studied, using the DDC formalism, the spontaneous excitation
of a two-level atom interacting with a conformally coupled massless
scalar field in the de Sitter-invariant vacuum in de Sitter
spacetime, and separately calculated the contributions of vacuum
fluctuations and radiation reaction to the rate of change of the
atomic energy. Both the case of a freely falling atom and that of a
static atom have been considered.

Remarkably, for a freely falling atom, we find a nonzero excitation
rate, which is exactly what one would obtain if there is  a bath of
thermal radiation at the Gibbons-Hawking temperature. We therefore
recover, in a different physical context, the results of Gibbons and
Hawking that reveals the thermal nature of de Sitter
spacetime~\cite{Gibbons}.

For a static atom, our results show that the atom also perceives a
thermal bath of radiation which is now a combined result of the
Gibbons-Hawking effect of de Sitter spacetime and the Unruh effect
associated with the inherent acceleration the atom must have in
order to be static. An interesting feature in contrast to the case
of a static atom outside a black hole~\cite{zwt yu2,zwt yu3} is that
the temperature of thermal bath as perceived by the static atom in
de Sitter spacetime is the  square root of the sum of the squared
Gibbons-Hawking temperature and the squared Unruh temperature
associated with the inherent acceleration of the atom.

Finally, It should be pointed out that we have only considered the
de Sitter invariant Bunch-Davies states, and it, therefore, remains
interesting to see whether the thermal response will still be
present when the vacuum states are replaced by a wider class of
states as those considered in
Refs.~\cite{Anderson2000,Anderson2005}.

\begin{acknowledgments}
 This work was supported in part by the National
Natural Science Foundation of China  under Grants No. 10575035 and
 No. 10775050, the SRFDP under
Grant No. 20070542002, and the Program for New Century Excellent
Talents in University under Grant No. 04-0784.
\end{acknowledgments}

\end{document}